\documentclass{article}[16pt]
\usepackage[utf8]{inputenc}
\usepackage{graphicx}
\begin{document}
%\begin{center}

\begin{center}
{\Large \bf Kinematic distances of galaxies in the Local Volume}

\bigskip

{\bf I.D.Karachentsev$^1$, A.A.Popova$^2$}

\bigskip

$^1$ Special Astrophysical Observatory,  Nizhnij Arkhyz,  Zelenchukskiy region,  Karachai-Cherkessian Republic, 369167, Russia \\
$^2$ Peter the Great St.Petersburg Polytechnic University, Saint Petersburg, 195251 Russia

{\bf Abstract.}
\end{center}

We consider the kinematic distances to nearby galaxies obtained by the Numerical Action Method (NAM) based on 
the Cosmic-flow-3 survey data. NAM distances are compared with 418 high-precision distances measured by the 
Tip of the Red Giant Branch (TRGB) method using the Hubble Space Telescope. We estimated the average difference 
$<D_{\rm NAM} -  D_{\rm TRGB}> = -0.30\pm0.08$ Mpc and the standard deviation of 1.57 Mpc. Approximately the 
same difference in the distance scale is obtained in comparison with less accurate distance estimates through 
the membership of galaxies in known groups or from the Tully-Fisher relation. We conclude that the NAM method 
provides distance estimates with an accuracy of 20\% within the Local Volume, which is valid for $\sim90$\% 
of the sky, except for the regions of the Virgo cluster and the Coma-I group.

\section{Introduction.}  Constructing a representative sample of nearby galaxies, limited by a fixed volume, 
is a necessary observational basis for cosmology on small scales. Efforts to create such a sample have been 
made repeatedly [1, 2]. Currently, the catalog of candidates for the population of the Local Volume (LV) 
contains about 1500 galaxies with expected distances $D$ within 11 Mpc [3]. The on-line version of this
Local Volume galaxy data base (LVGDB) [4], supplemented by recently discovered objects, is available at 
http://www.sao.ru/lv/lvgdb.

With an ideal unperturbed Hubble flow with a Hubble parameter $H_0 = 73$ km~s$^{-1}$ Mpc$^{-1}$, galaxies 
with radial velocities relative to the centroid of the Local Group $V_{\rm LG} < 800$ km~s$^{-1}$ would fall 
into the designated LV. The presence of inhomogeneities in the distribution of matter introduces anisotropy 
into the local Hubble flow. According to [5], our Galaxy and its neighbors are participating in the motion 
towards the nearby Virgo cluster ($D = 16.5$ Mpc) with an amplitude of $\sim180$ km~s$^{-1}$ and in the 
expansion of the Local Void with a characteristic velocity of $\sim260$ km~s$^{-1}$. The directions of these 
local flows are approximately mutually perpendicular.

To construct a more detailed map of the local field of radial velocities, the Numerical Action Method = NAM 
[6, 7] was proposed, which takes into account the location and masses of nearby attractors (groups of 
galaxies), as well as the most significant neighboring clusters and voids outside the LV. Kourkchi et al. 
[8] developed a convenient scheme for determining the kinematic (NAM) distances of LV galaxies from their 
coordinates and radial velocity $V_g$ relative to the center of our Galaxy. In this case, estimates of the 
distance of galaxies made by various methods were used. The main array consisted of high-precision measurements 
of distances by the luminosity of the Tip of the Red Giant Branch (TRGB) with a total number of about 400. 
The accuracy of this method reaches $\sim5$\%. Actually, the radius of the LV, 11 Mpc, was determined just 
by the ability to measure the TRGB-distance of a galaxy from its images obtained with the Hubble Space Telescope (HST) during 
one orbital period. In addition to this universal method, applicable to galaxies of any morphological type, 
a small number of distance estimates were used from Cepheids, supernovae, and surface brightness fluctuations.

It should be noted that measuring the distances of galaxies with an accuracy of $\sim5$\% is a laborious, 
expensive procedure, unlike measuring radial velocities. Mass surveys of radial velocities in the optical and 
radio ranges [9, 10 11, 12] have significantly enriched our data on the field of radial velocities of galaxies 
in the LV. This growth continues with the introduction of ever larger telescopes into observations. On the other 
hand, episodic programs for measuring TRGB-distances on the aging HST make an increasingly smaller contribution 
to the overall panorama of galaxy distances beyond $D \sim3$ Mpc. Therefore, estimates of kinematic (NAM) 
distances, which allow us to determine the absolute luminosity and other important parameters of nearby galaxies, 
are becoming more relevant.

The purpose of this work is to estimate the accuracy of kinematic distances of galaxies in the Local Volume, 
using observational data from the LVGDB (http://www.sao.ru/lv/lvgdb).

\section{Observational data.}  The LVGDB contains TRGB-distance values for 473 LV galaxies with 
measured radial velocities. We excluded from them 43 members of the Local Group with distances $D < 1$ Mpc, 
whose virial motions are not related to the general pattern of the peculiar velocity field of the LV. For 
the remaining 430 galaxies, kinematic NAM-distances were determined according to the diagram [8]. For most 
of these galaxies, their radial velocities are measured with an error of less than 8 km~s$^{-1}$, 
corresponding to an error in $D_{\rm NAM}$ of less than 0.1 Mpc, which we neglected. A machine-readable list of 
these galaxies with $D_{\rm TRGB}$ and $D_{NAM}$ values can be provided upon individual request.
\begin{figure}
\includegraphics[scale=0.5]{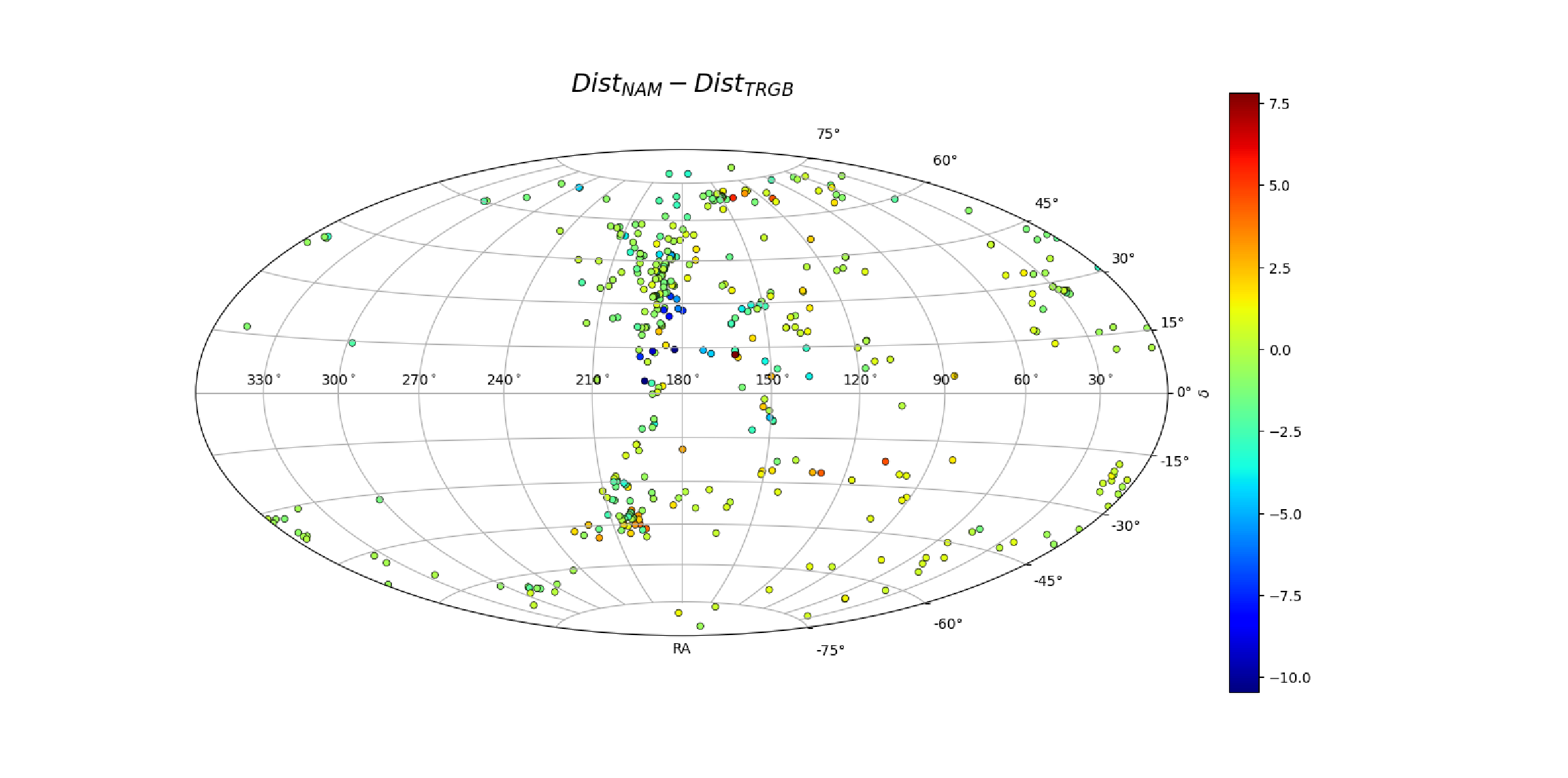}
\includegraphics[scale=0.5]{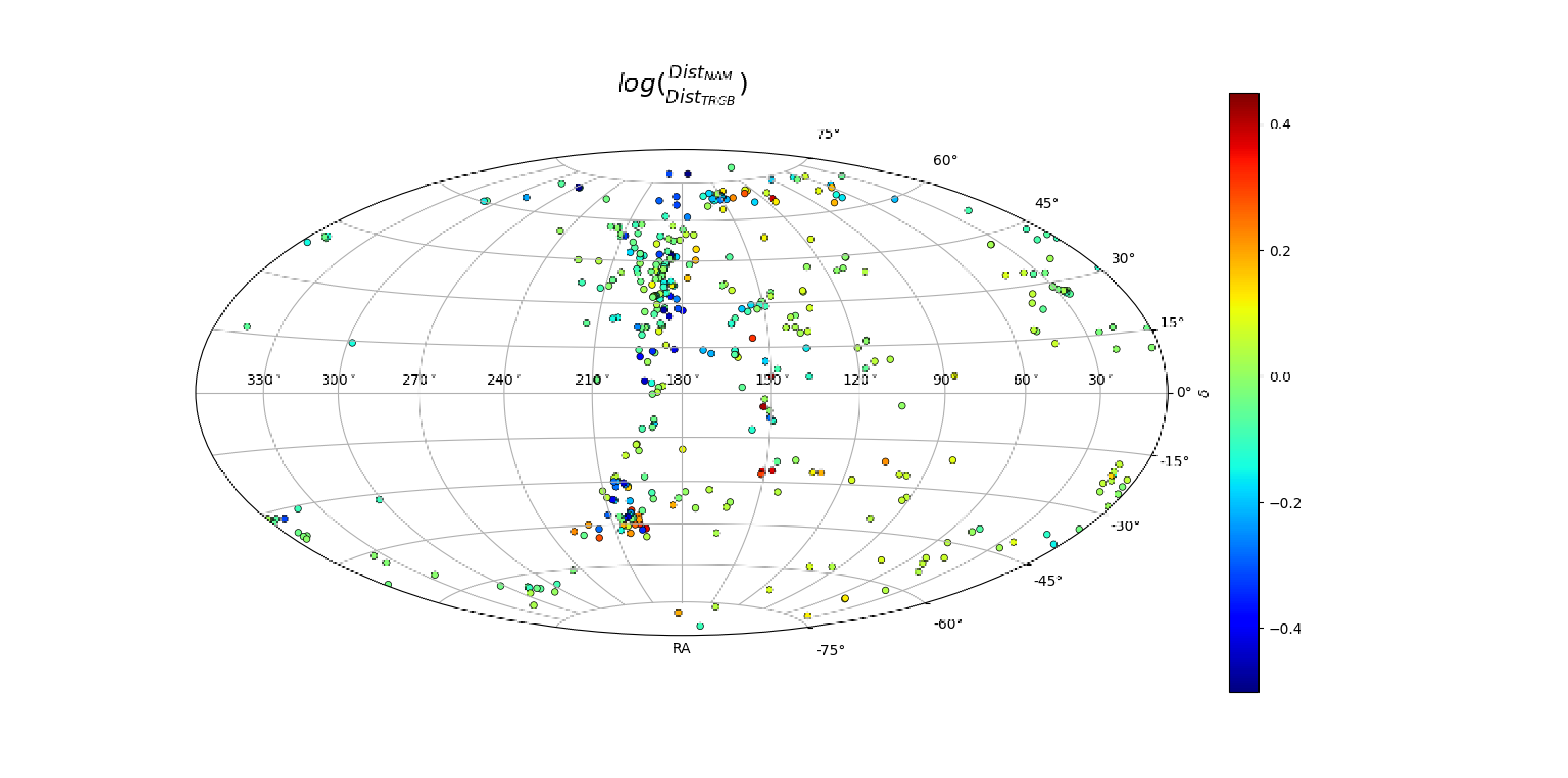}
\caption{Distribution of galaxies in the Local Volume by the difference (top) and ratio (bottom) of the 
distances $D_{\rm NAM}$ and $D_{\rm TRGB}$.}
\end{figure}

\section{The local kinematic distances.} The top panel of Fig.1 reproduces the distribution of 430 LV 
galaxies according to the difference in distance estimates $\Delta = D_{\rm NAM} - D_{\rm TRGB}$ in 
equatorial coordinates. The color scale on the right reflects the magnitude of $\Delta$ in Mpc. Over the 
predominant area of the sky, galaxies, denoted by circles, have a yellow-green color, corresponding to 
the difference in distance estimates within 1--2 Mpc. The left side of the sky map is characterized by 
a vast emptiness, due to the fact that the Local Void extends almost to the border of the Local Group of 
galaxies.

The bottom panel of Fig.1 shows the distribution on the sky of LV galaxies according to the magnitude of 
the distance ratio $D_{\rm NAM}/D_{\rm TRGB}$ in the same coordinates. This diagram complements the 
previous one, since the error $\sigma(D_{\rm TRGB}) \sim5$\% affects the value of ($D_{\rm NAM} - 
D_{\rm TRGB}$) differently for nearby and distant galaxies. Typical virial velocities in groups, 
$\sigma_v \sim (70-100)$ km~s$^{-1}$, are also the reason for the scatter of galaxies in terms of the 
value of $D_{\rm NAM}/D_{\rm TRGB}$, especially in nearby groups around M81 and CenA=NGC5128.

The distribution of the number of LV galaxies according to the magnitude of the difference in distance 
estimates $\Delta = D_{\rm NAM} - D_{\rm TRGB}$ is shown in Fig.2. We excluded 12 strongly deviating 
galaxies with $\mid\Delta\mid > 5.0$, their relative number does not exceed 3\%. The histogram $N(\Delta)$ 
fits well with the Gaussian function having parameters $<\Delta> = -0.3\pm0.08$ Mpc and $\sigma_{\Delta} = 
1.57$ Mpc.

\begin{figure}
\includegraphics[height=8cm] {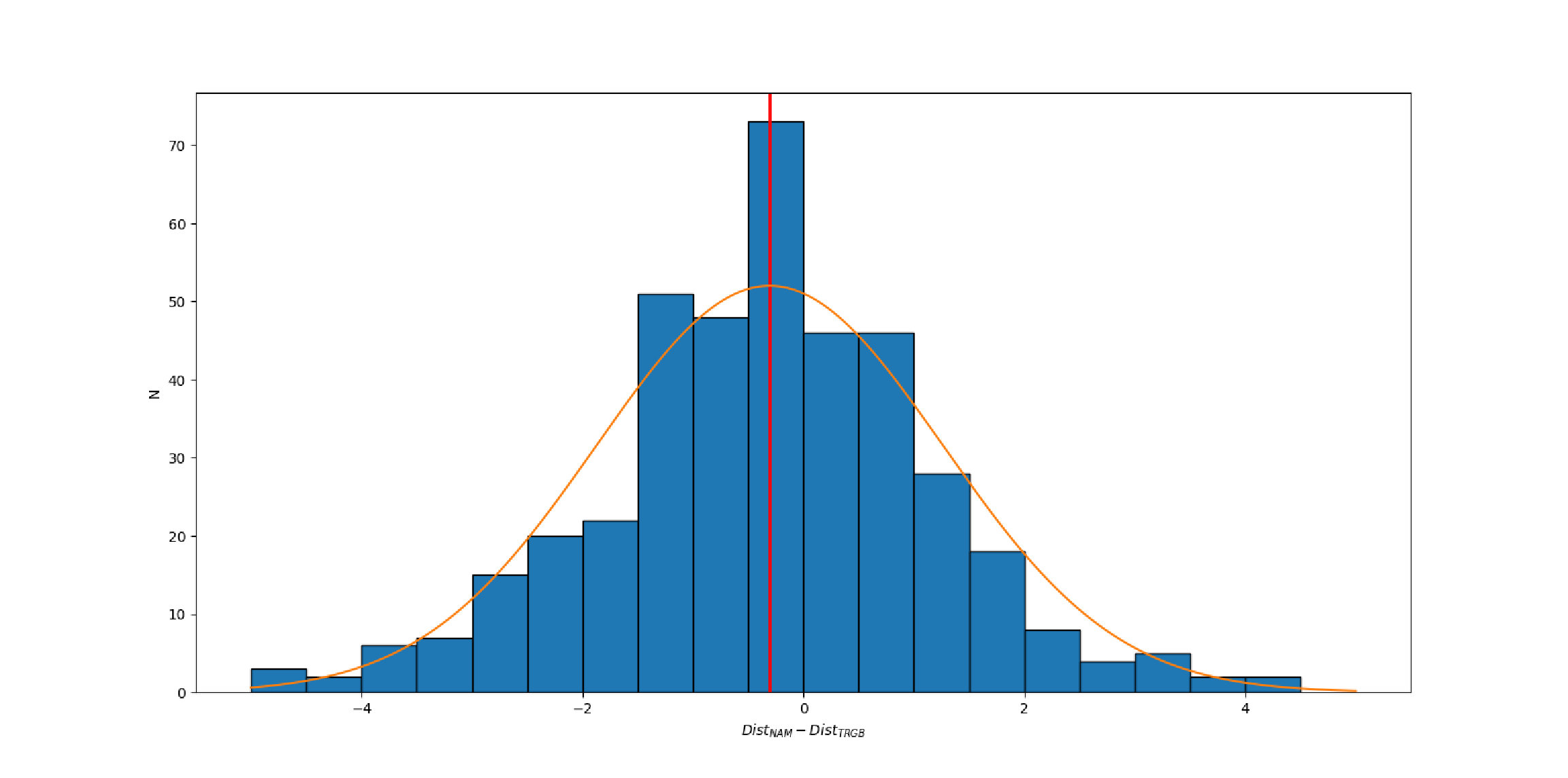}
\caption{Distribution of number of galaxies in the Local Volume by the magnitude of difference 
in their distance estimates $\Delta = D_{\rm NAM} - D_{\rm TRGB}$.}
  \end{figure}

\begin{table}
\caption{A list of 12 ``TRGB--NAM'' outliers}  
\begin{tabular}{lcccccc} \hline

    Name      &  RA       &   Dec  &      $V_g$   & $D_{\rm TRGB}$ & $D_{\rm NAM}$  &  $\Delta_D$  \\
\hline
              &    deg    &    deg &   km~s$^{-1}$&  Mpc    &    Mpc    &    Mpc  \\
\hline
  UGC04998    &  141,30   &  68,38 &     742      & 8,24    &   13,69   &    5,45 \\
  NGC3379     &  161,96   &  12,58 &     818      &11,32    &   19,13   &    7,81  \\ 
  BTS76       &  179,68   &  27,58 &     436      &12,59    &   5,17    &   -7,42 \\  
  LV J1205+28 &  181,39   &  28,23 &     491      &11,53    &   5,65    &   -5,88  \\ 
  LV J1207+31 &  181,96   &  31,55 &     559      &12,02    &   6,64    &   -5,38  \\
  IC3023      &  182,51   &  14,37 &     755      &17,0     &   6,73    &  -10,27  \\ 
  LV J1217+32 &  184,38   &  32,53 &     456      &12,59    &   5,32    &   -7,27 \\ 
  AGC229053   &  184,56   &  25,57 &     409      &12,47    &   4,45    &   -8,02 \\ 
  IC3341      &  186,60   &  27,75 &     368      &11,64    &   3,94    &   -7,70 \\  
  KDG177      &  189,99   &  13,78 &     958      &17,0     &   7,74    &   -9,26 \\   
  UGC07983    &  192,45   &  03,84 &     619      &16,52    &   6,06    &  -10,46 \\ 
  UGC08061    &  194,18   &  11,93 &     520      &12,59    &   5,21    &   -7,38 \\ \hline 
 \end{tabular}
 \end{table}

Table 1 presents a list of 12 ``outlier'' galaxies with $\mid\Delta\mid > 5$ Mpc. Its first column lists 
the names of galaxies as they are designated in LVGB; columns (2, 3) give the equatorial coordinates 
of the galaxies; column (4) contains radial velocities relative to the center of the Galaxy; the 
distances of galaxies and their difference in Mpc are given in the last three columns. Analysis of 
these data reveals several reasons for the large deviations $\Delta$. For the galaxy UGC4998, an 
erroneous determination of the $D_{\rm TRGB}$ distance was made on the ``color-magnitude'' diagram 
[13] due to confusion between the $RGB$ and $AGB$ sequence stars. The galaxies IC3023, KDG177, UGC7983, 
and UGC8061 are located inside the virial zone of the Virgo cluster, where NAM-distance estimates are 
distorted by virial motions. The galaxies BTS76, LVJ1205+28, LVJ1207+31, LVJ1217+32, AGC229053, and 
IC3341 belong to members of the specific Coma I group, around NGC4278 [14], which is located at the 
border of the zero-velocity radius of the Virgo cluster, $R_0\simeq23^{\circ}$. The galaxies of this 
group have negative peculiar velocities $V_{\rm pec} \sim -800$ km~s$^{-1}$. The nature of this anomaly 
remains a mystery.

At an average distance of LV galaxies of 8.0 Mpc and a standard deviation $\sigma_{\Delta}$ = 1.57 Mpc, 
the relative error in estimating the NAM distance is 19\%. Here we made a quadratic subtraction of the 
relative error of 5\% caused by errors in measuring TRGB distances.

For dwarf galaxies in the Local Volume groups, virial velocities of $\sim80$ km~s$^{-1}$ introduce a 
relative error in the NAM distance estimate of about 12\%. Therefore, when using the average radial 
velocity in a sufficiently populated group, one can expect a typical error in the average NAM distance 
of $\sim15$\%.

\begin{table}
\caption{The major Local Volume  galaxies.}
\begin{tabular}{lccc}   \hline
 Name      &  $D_{\rm TRGB}$ &    $D_{\rm NAM}$  &     $\Delta_D$\\
 \hline

NGC 253    &  3.70     &     3.83    &     0.13
 \\
 NGC 628   &   10.19   &      9.58   &     -0.61
 \\
 NGC 891   &    9.95   &      10.39  &      0.44
 \\
 NGC1291   &    9.08   &       9.16  &       0.08
 \\
 IC 342    &   3.28    &      3.70   &      0.42
\\
 NGC2683   &   9.82    &    10.51    &     0.69
 \\
 NGC2903   &   9.15    &    10.86    &     1.71
 \\
 M 81      &   3.70    &      1.84   &     -1.86
\\
 NGC3115   &  9.68     &    10.39    &     0.71
 \\
 NGC3184   & 11.12     &   11.42     &    0.30
  \\
 NGC4258   &  7.66     &     7.33    &     -0.33
\\
 NGC4594   &  9.55     &    8.71     &     -0.84
 \\
 NGC4736   &  4.41     &    4.03     &     -0.38
\\
 NGC5055   &  9.04     &    7.18     &     -1.86
 \\
 NGC5128   &  3.68     &    3.46     &     -0.22
 \\
 M 51      &  8.40     &    7.13     &     -1.27
 \\
 NGC5236   &  4.90     &    2.97     &     -1.93
 \\
 M 101     &  6.95     &    4.85     &     -2.10
 \\
 NGC6744   &  9.51     &    8.63     &     -0.88
\\
 NGC6946   &  7.73     &    5.41     &     -2.32
\\
\hline
 NGC3379   & 11.32     &   19.13     &    7.81
  \\
 NGC3627   & 11.12     &    6.72     &    -4.40
 \\
\hline
\end{tabular}
\end{table}

To test this assumption, we listed in Table 2 all 22 major LV galaxies with the Milky Way - like luminosity 
and accurate distance estimates.These galaxies dominate in mass in their groups, and it can be expected that their 
peculiar velocities are small. As follows from these data, the average difference in distance estimates 
for them is $<D_{\rm NAM} - D_{\rm TRGB}> = -0.51\pm0.27$ Mpc, and the relative standard deviation of the 
difference is 15\%. Note that the following galaxies were added to the 16 galaxies in the table with TRGB 
distances: IC342 with a distance estimate from Cepheids, NG3115 with a distance estimate from surface 
brightness fluctuations, and NGC3184, M51, whose distances are determined by the luminosity of 
supernovae. The bottom two lines of Table 2 show the galaxies NGC3379 and NGC3627, which are massive but 
not dominant in their groups in the Leo Spur region [15]. According to [7], this region has a complex 
structure of the peculiar velocity field due to the projection onto the line of sight of two structures: 
the Leo Spur and the Leo cloud [16] and the presence of a void behind the Leo cloud.

\section{Comparison with other distance estimates.} The LVGDB database contains 94 galaxies with measured 
radial velocities, whose distances are estimated by the assumed membership (mem) in nearby groups. Of these, 
we excluded 9 galaxies with deviations \\
$\mid D_{\rm NAM} -D_{\rm mem}\mid > 5.0$ Mpc. They are located in the Leo I (NGC3379) and Sombrero 
(NGC4594) groups with large virial velocities. For the remaining 85 galaxies, we obtained the values 
$<D_{\rm NAM} - D_{\rm mem}> = -0.23$ Mpc and 
$\sigma_{\Delta}$ = 1.97 Mpc. The relative error of the difference $\Delta$ for them is 21\%. Assuming 
that the relative error of NAM distances is 15\%, we find approximately the same value for the error of 
mem-distances, 15\%.

The LVGDB database also contains 159 galaxies with distance estimates based on the Tully-Fisher method [17], 
using the relationship between the luminosity of a galaxy and the width of the 21-cm HI line. We excluded 
from consideration 16 galaxies with a large difference $\mid D_{\rm NAM} - D_{\rm TF}\mid > 16$ Mpc, 
which make up 10\% of this sample. All of them, except for one, are located inside the virial zone of 
the Virgo cluster ($n = 9$) or the Coma I group ($n = 6$) with anomalous peculiar velocities. Among these 
excluded objects, the galaxy UGC7774 (RA = 189.03$^{\circ}$, Dec = +40.00$^{\circ}$) has estimates 
$D_{\rm TF} = 22.6$ Mpc and $D_{\rm NAM} = 7.2$ Mpc, but is not in any known group. For the remaining 
143 galaxies, we obtained the values $<D_{\rm NAM} - D_{\rm TF}> = -0.76\pm0.35$ Mpc and 
$\sigma(D_{\rm NAM} - D_{\rm TF}) = 4.19$ Mpc. Given the average distance of the galaxies in this sample, 
$<D_{\rm TF}> = 11.1$ Mpc, we obtain a relative error in the TF distance estimate of 31\%. This value is 
noticeably larger than the typical error of the TF method, 20\%, obtained for spiral galaxies. This 
difference is quite understandable, since dwarf galaxies dominate in the LV, and their irregular shape 
makes the correction for the inclination of their axis of rotation to the line of sight uncertain.

\section{Concluding remark.} A comparison of kinematic distance estimates made in the NAM model with 
distance measurements using the TRGB method and other methods shows a small but systematic shift in the 
NAM scale, $<D_{\rm NAM} - D_{\rm TRGB}> = -0.30\pm0.08$ Mpc. The relative error in determining the 
kinematic distance is $\sim20$\% for individual galaxies in the Local Volume. If there are several members 
in the group with measured radial velocities, the error of the NAM method can be improved to 15\%. This 
makes the NAM method more preferable compared to the Tully-Fisher method, the relative error of which for 
the LV population we estimated as 31\%.

However, there are two adjacent areas in the sky: the zone of galaxies falling onto the Virgo cluster 
with a radius of $\sim23^{\circ}$ and the zone of large negative peculiar velocities in the Coma I group 
north of the Virgo cluster, where the kinematic method gives large errors and is practically inapplicable. 
The total area of these anomalous regions occupies only 10\% of the sky. Taking into account this caveat, the 
kinematic method in the NAM model is quite suitable for the mass determination of distances for nearby 
galaxies with an error of 15-20\%.

The study was carried out with the support of a grant from the Russian Science Foundation No. 24-12-00277.

\bigskip

{\bf REFERENCES}

1. R.C. Kraan-Korteweg, G.A.Tammann,  Astron. Nachr., {\bf 300}, 181, 1979.

2. I.D.Karachentsev, V.E.Karachentseva, W.K.Huchtmeier, D.I.Makarov,  Astron. J., {\bf 127}, 2031, 2004.

3. I.D.Karachentsev,  D.I.Makarov,   E.I.Kaisina,   Astron. J., {\bf 145}, 101, 2013.

4. E.I.Kaisina, D.I.Makarov, I.D.Karachentsev, S.S.Kaisin,  Astrophys. Bull., {\bf 67}, 115, 2012.

5. R.B.Tully, E.J.Shaya, I.D.Karachentsev et al.,  Astrophys. J., {\bf 676}, 184, 2008.

6. P. J. E.Peebles,   S.D.Phelps,  E.J.Shaya,  R.B.Tully,  Astrophys. J., {\bf 554}, 104,  2001.

7. E.J.Shaya,  R.B.Tully,  Y.Hoffman,  D.Pomar\'{e}de, Astrophys. J., {\bf 850}, 207,  2017.

8. E.Kourkchi,  H.M.Courtois, R.Graziani et al., Astron. J., {\bf 159}, 67, 2020.

9.  K.N.Abazajian, J.K.Adelman-McCarthy,  M.A.Ag\"{u}eros et al., Astrophys. J. Suppl., {\bf 182}, 543, 2009.

10. B.S.Koribalski,  L.Staveley-Smith, V.A.Kilborn et al., Astron. J., {\bf 128}, 16, 2004.

11.  M.P.Haynes,  R.Giovanelli, A.M.Martin et al.,  Astron. J., {\bf 142}, 170, 2011.

12. C.-P.Zhang,  M.Zhu,  P.Jiang et al.,   Science China Physics, Mechanics, and Astronomy, {\bf 67}, 219511, 2024.

13.  J.Alonso-Garcia, M.Mateo, A.Aparicio,  Publ.  Astron. Soc. Pac.,  {\bf 118}, 580, 2006.

14. I.D.Karachentsev,  O.G.Nasonova,  H.M.Courtois,   Astrophys. J., {\bf 743}, 123, 2011.

15.I.D. Karachentsev, E.I.Kaisina, V.E.Karachentseva,  Mon.  Not.  Roy.  Astron. Soc., {\bf 521}, 840, 2023.

16. R.B.Tully,  Nearby Galaxies Catalog (Cambridge: Cambridge University Press), 1988.

17. R.B.Tully,  J.R.Fisher, Astron. and Astrophys., {\bf 54}, 661,  1977.

 \end{document}